\documentclass{pspum-l}
\usepackage{amssymb}

\usepackage{graphicx}

\usepackage[cmtip,all]{xy}

\usepackage{rotating}

\usepackage{bbm}

\copyrightinfo{2015}{J\"urgen Fuchs and Christoph Schweigert}

\theoremstyle{definition}

\theoremstyle{remark}

\def\BrPic  {\mathrm{BrPic}}
\def\Bun    {\mathrm{Bun}}
\def\cala   {{\mathcal A}}
\def\calc   {{\mathcal C}}
\def\calm   {{\mathcal M}}
\def\calw   {{\mathcal W}}
\def\calz   {{\mathcal Z}}
\def\cobord {\mathrm{cobord}}
\def\tft    {\mathrm{tft}}
\def\Vect   {\mathrm{Vect}}

\begin{document}

\title[Symmetries and defects in 3d topological field theory]
{Symmetries and defects \\ in three-dimensional topological field theory}

\author{J\"urgen Fuchs}
\address{Teoretisk fysik, Karlstads universitet, Universitetsgatan 21,
651\,88 Karlstad, Sweden}
\email{juergen.fuchs@kau.se}
\thanks{JF is supported by VR under project no.\ 621-2013-4207.}

\author{Christoph Schweigert}
\address{Algebra und Zahlentheorie, Fachbereich Mathematik, 
Universit\"at Hamburg, Bundesstra\ss{}e 55, 
20148 Hamburg, Germany}
\email{christoph.schweigert@uni-hamburg.de}
\thanks{CS is partially supported by the Collaborative Research Centre 676 ``Particles, 
Strings and the Early Universe - the Structure of Matter and Space-Time'', by the RTG 1670
``Mathematics inspired by String theory and Quantum Field Theory'' and by the DFG Priority 
Programme 1388 ``Representation Theory''. \\
We are grateful to the Erwin-Schr\"odinger-Institute (ESI) for the
hospitality during the programs ``Modern trends in topological field theory'' 
and ``Topological Phases of Quantum Matter''
while part of this work was done.}


\date{January 15, 2015}

\begin{abstract}
Boundary conditions and defects of any codimension are natural parts of any quantum field theory. 
Surface defects in three-dimensional topological field theories of Turaev-Reshetikhin type 
have applications to two-dimensional conformal field theories, in solid state physics and 
in quantum computing. We explain an obstruction to the existence of surface defects that 
takes values in a Witt group. We then turn to surface defects in Dijkgraaf-Witten theories 
and their construction in terms of relative bundles; this allows one to exhibit 
Brauer-Picard groups as symmetry groups of three-dimensional topological field theories.
\end{abstract}

\maketitle

\section{Introduction}

In this contribution we discuss the relation between two important structures in 
quantum field theory: symmetries on the one hand, and defects of various codimensions
on the other. While it is hardly necessary to emphasize the importance of symmetries, 
defects in quantum field theories or, in a different formulation, quantum field theories 
on stratified spaces, and their relation to symmetries are a more recent 
topic of interest. 

Our exposition is organized as follows. We first discuss a few general aspects of 
topological codimension-1 defects, both in two-dimensional and in three-di\-men\-sional
quantum field theories. A special emphasis is put on the relation between
\emph{invertible} topological defects and symmetries.
Afterwards we consider defects for a particular class of three-dimensional topological
field theories: Dijkgraaf-Witten theories. An advantage of these models is
that they admit a ma\-the\-matically precise formulation as gauge theories, see 
\cite{DW90,Fr95,Mo14} and references therein.
Apart from providing insights in a language that is close to standard field theoretical 
formulations, this allows one to uncover interesting relations between categories 
of (relative) bundles and recent results in representation theory.


\section{Topological defects in quantum field theories}

It has been recently realized in several different contexts that boundary conditions 
and defects of various codimensions constitute important parts of the structure 
of a quantum field theory. (Alternatively, one might say that it is instructive 
to consider the quantum field theory not only on smooth manifolds, but also on stratified spaces.)
Codimension-1 defects, also called interfaces, 
separate regions that can support different quantum field theories.
Such defects arise naturally in applications, ranging 
from condensed matter systems, where they appear as interfaces between different 
phases of matter, to domain walls in cosmology. The aspect we wish to emphasize in 
this note is the fact that such defects provide a lot of structural insight into 
quantum field theories, including in particular their symmetry structures.

Similarly as in the case of boundary conditions, there are various different types of 
defects one can consider. In particular, one can impose various conservation conditions 
of physical quantities on defects. An interesting subclass of codimension-1 defects 
are {\em topological} defects. They are characterized by the property that the 
values of correlators for configurations with defects do not change when the 
location of a defect is only slightly changed, that is, changed by a 
homotopy without crossing any field insertions or other defects. Examples of such
defects have been known for a long time (see e.g.\ \cite{DW82}). Consider, for
instance, the two-dimensional Ising model, defined by a ${\mathbb Z}_2$-valued 
variable on the vertices of a two-dimensional CW-com\-plex. 
Select a line that crosses bonds transversally, and change the coupling on each bond 
that is crossed by the line from ferromagnetic to anti-ferromagnetic. In the
continuum limit this provides a topological defect line in the critical Ising model.


\subsection{Symmetries from invertible topological defects}

A particularly important subclass of topological defects are the 
{\em invertible} topological defects. Owing to their topological nature, two 
topological defects can be brought to coincidence, leading to a fusion product 
of defects. The precise mathematical formulation of the relevant monoidal 
structure depends on the dimension in which the quantum field theory is
defined. For the moment, let us consider two-dimensional quantum field theories. 
All statements made in the sequel have the status of theorems \cite{FFRS04,FFRS07} 
in the case of two-dimensional rational conformal field theories; the general 
picture should be much more widely applicable, though. For a two-dimensional 
rational conformal field theory, the topological line defects form a monoidal 
category, with morphisms provided by field insertions which can change the 
type of defect. In particular, there is an ``invisible defect'', which is
a monoidal unit $\mathbf 1$ in the category of defects.

An \emph{invertible} defect $D$ is characterized by the fact that there 
exists another defect $D^\vee$ with the property that the fusion of $D$ and 
$D^\vee$ gives the invisible defect, 
  $$
  D\otimes D^\vee \cong \mathbf 1 \cong D^\vee \otimes D \,.
  $$
(Thus in particular for every theory the invisible defect is invertible.)
This behaviour leads to the relation
  $$
  \begin{picture}(210,90)
  \put(20,0)   {\includegraphics{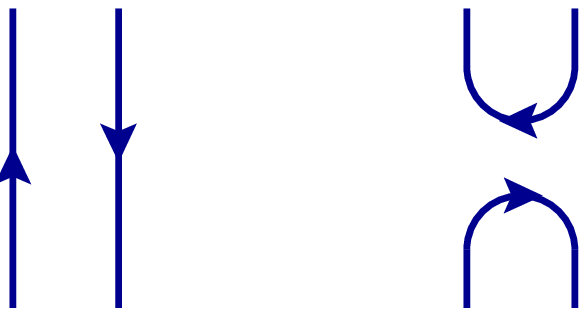}}
  \put(13,28)  {\scriptsize$ D $}
  \put(57.6,28){\scriptsize$ D^\vee $}
  \put(82,42)  {$ = $ }
  \put(101,42) {$ \mathrm{dim}(D)^{-1} $ }
  \put(155,0){
  \put(34,20)  {\scriptsize$ D^\vee $}
  \put(33.3,65){\scriptsize$D ^\vee $}
  }
  \end{picture}
  $$
where $\mathrm{dim}(D)$ is the (quantum) dimension of the defect, which
for an invertible defect can take the values $+1$ 
(in unitary theories this is the only possibility) or $-1$. This relation is to be 
understood as an identity of correlators when applied locally in any 
configuration of fields and defect lines. With this relation, it is immediate
to deduce a connection with symmetries: one has equalities of the form
  $$
  \begin{picture}(360,57)
  \put(0,0)    {\includegraphics{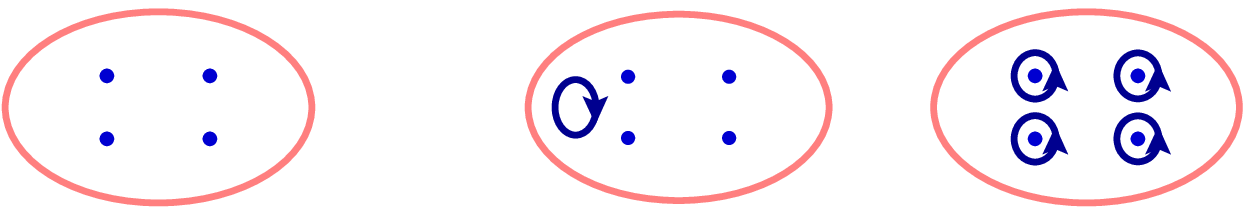}}
  \put(98,27) {$ =\, \displaystyle\frac1{\dim(D)} $ }
  \put(251,27) {$ = $ }
  \end{picture}
  $$
between correlators of different bulk fields.

Let us discuss this issue explicitly for the case of the two-dimensional critical 
Ising model. There are then three primary fields, commonly referred to as the
identity field $1$, the spin field $\sigma$ and the energy field $\epsilon$. Their 
conformal weights are $0$, $\frac1{16}$ and $\frac12$,
respectively. Indecomposable topological line defects of the critical Ising model 
turn out to be in bijection with these fields: in particular, $1$ corresponds to 
the invisible defect and $\epsilon$ to an invertible defect. (It is in fact 
the latter defect that amounts to changing the couplings from ferromagnetic to 
antiferromagnetic.) The action of this specific invertible defect on bulk fields 
is as follows: it leaves the bulk fields corresponding to $1$ and to $\epsilon$ 
invariant and changes the sign of $\sigma$. Inside RCFT correlators for the 
critical Ising model we thus have the equalities
  $$
  \begin{picture}(350,32)
  \put(0,2){
  \put(0,0)    {\includegraphics{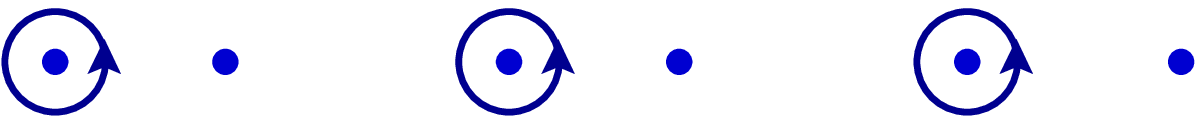}}
  \put(13.8,5.5){\scriptsize$ 1 $ }
  \put(29,25.9) {\scriptsize$ \epsilon $ }
  \put(44,13.5) {$ = $ }
  \put(62.7,1.9){$ 1 $ }
  }
  \put(130,2){
  \put(15.2,6.9){\scriptsize$ \epsilon $ }
  \put(29.3,25.9) {\scriptsize$ \epsilon $ }
  \put(44,13.5) {$ = $ }
  \put(64.0,4.7){$ \epsilon $ }
  }
  \put(263,2){
  \put(13.8,7.4){\scriptsize$ \sigma $ }
  \put(29.7,25.9) {\scriptsize$ \epsilon $ }
  \put(44,13.5) {$ = ~~ - $ }
  \put(75.4,5.5){$ \sigma $ }
  }
  \end{picture}
  $$

As can be seen from these pictures, there is a natural action of invertible topological 
line defects on field insertions: wrapping such a defect around a bulk field insertion
yields another bulk field.

A further advantage of having a realization of symmetries in terms of invertible topological
defects is seen by studying what happens when the defect is moved to a boundary with 
boundary condition $M$: this process yields another boundary condition $M'$, according to
  $$
  \begin{picture}(201,88)
  \put(0,-1){
  \put(27,0)     {\includegraphics{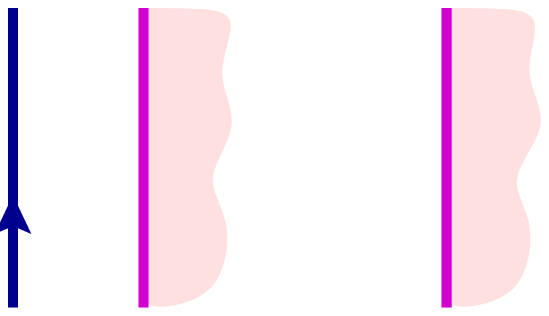}}
  \put(19.4,16)  {\scriptsize$ D $}
  \put(55.9,18)  {\scriptsize$ M $}
  \put(119,40)   {$ = $}
  \put(141.2,18) {\scriptsize$ M' $}
  }
  \end{picture}
  $$

In a similar manner, by wrapping around various structures in an obvious way,
the group of invertible topological defects 
acts as a symmetry group on {\em all} data of the field theory, including field insertions
of bulk, boundary and disorder fields, boundary conditions and types of defects. 
In the framework of the TFT construction of correlators of two-dimensional
rational conformal field theories \cite{SFR06}, isomorphism classes of invertible topological
defects can be explicitly classified \cite{FFRS04,FFRS07}. In the example of the critical Ising
model one finds a symmetry group ${\mathbb Z}_2$, consisting of the invisible defect
and the $\epsilon$-defect described above, while e.g.\ for the critical three-state Potts model 
one obtains a non-abelian symmetry group $S_3$. 

For the present purposes, it is important to realize that topological codimension-1 defects can 
also exist in more general classes of quantum field theories. Moreover, in the general case
geometric considerations suggest a natural action of such defects on field theoretic structures,
like boundary conditions and defects of various codimensions, as well.
In short: defects can wrap around field theoretic structures.


\subsection{T-dualities and Kramers-Wannier dualities from topological line defects}

Before turning to higher-dimensional field theories, we point out that topological line defects are
indeed of much wider use: they can also implement Kramers-Wannier dualities and T-dualities.
If a general topological defect wraps around a bulk field, the following situation is created:
  $$
  \begin{picture}(190,67)
  \put(0,-3){
  \put(0,0)  {\includegraphics{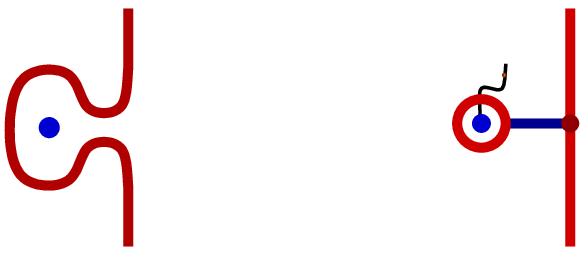}}
  \put(11.4,27.2){\scriptsize$ \phi $ }
  \put(39.2,58){\scriptsize$ D $ }
  \put(55,35){ $\displaystyle = ~
                \underset{{\rm defects}~D_i}{\underset{{\rm intermediate}}{\sum}}$ }
  \put(143.7,57.7){\scriptsize$ \phi $ }
  \put(131,23.3){\scriptsize$ D $ }
  \put(149.6,30.1){\scriptsize$ D_i $ }
  \put(166.9,58){\scriptsize$ D $ }
  }
  \end{picture}
  $$

In this way a bulk field $\phi$ is turned by the defect $D$ into a disorder field. To obtain 
an order-disorder duality, one also needs the opposite process, turning disorder fields 
into ordinary local bulk fields. It can be shown \cite{FFRS04,FFRS07} that to this
end the dual defect $D^\vee$ must be used, and that in this case one turns the disorder field 
back into a bulk field if and only if the fused defect $D\,{\otimes}\, D^\vee$ is a direct sum 
of invertible defects. This condition can be examined in concrete models.
It is in particular satisfied for the defect corresponding to the spin field $\sigma$ 
in the critical Ising model, thanks to the well known fusion rule 
$\sigma \,{\otimes}\, \sigma \,{\cong}\, 1 \,{\oplus}\, \epsilon$; 
this defect indeed produces the action of the Kramers-Wannier duality in the critical model. 

Again we have a natural action on correlators:
  $$
  \begin{picture}(400,137)
  \put(0,75)   {\includegraphics{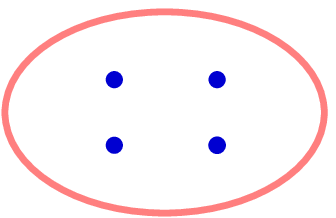}}
  \put(163,75) {\includegraphics{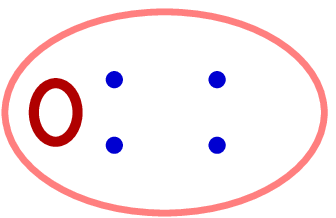}}
  \put(138,0)  {\includegraphics{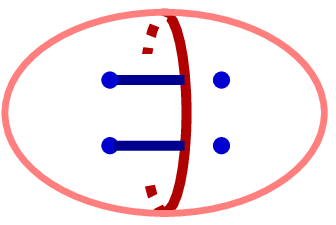}}
  \put(264,0)  {\includegraphics{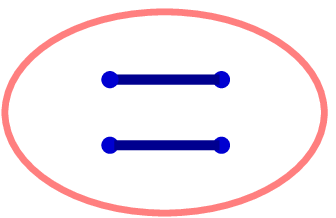}}
  \put(106,102){$\displaystyle = ~ \frac1{\dim(D)} $}
  \put(81,27)  {$\displaystyle = ~ \frac1{\dim(D)} $}
  \put(243,27) {$ = $}
  \end{picture}
  $$

\noindent
which shows how correlators involving only bulk fields are related to correlators involving
disorder fields. Specifically, through the action of the $\sigma$-defect the correlator of four 
spin fields on a sphere can be seen to be equal to the correlator of four disorder fields, according to
  $$
  \begin{picture}(400,137)
  \put(0,75)   {\includegraphics{pic_ffrs3_4I.eps}}
  \put(147,75) {\includegraphics{pic_ffrs3_4J.eps}}
  \put(138,0)  {\includegraphics{pic_ffrs3_4K.eps}}
  \put(264,0)  {\includegraphics{pic_ffrs3_4L.eps}}
  \put(106,102){$\displaystyle = ~ \frac1{\sqrt2} $}
  \put(97,27)  {$\displaystyle = ~ \frac1{\sqrt2} $}
  \put(243,27) {$ = $}
  \put(4,88) {
  \put(26.7,.2)  {\scriptsize $ \sigma $}
  \put(26.9,31.4){\scriptsize $ \sigma $}
  \put(56.3,.2)  {\scriptsize $ \sigma $}
  \put(56.5,31.4){\scriptsize $ \sigma $}
  }
  \put(151,88) {
  \put(17.5,26)  {\scriptsize $ \sigma $}
  \put(26.7,.2)  {\scriptsize $ \sigma $}
  \put(26.9,31.4){\scriptsize $ \sigma $}
  \put(56.3,.2)  {\scriptsize $ \sigma $}
  \put(56.5,31.4){\scriptsize $ \sigma $}
  }
  \put(142,14) {
  \put(25.0,1)   {\scriptsize $ \mu $}
  \put(25.3,31.9){\scriptsize $ \mu $}
  \put(37,11.9)  {\scriptsize $ \epsilon $}
  \put(37,21.3)  {\scriptsize $ \epsilon $}
  \put(51.9,16)  {\scriptsize $ \sigma $}
  \put(57.3,1)   {\scriptsize $ \sigma $}
  \put(57.5,31.9){\scriptsize $ \sigma $}
  }
  \put(267,14) {
  \put(26.0,0)   {\scriptsize $ \mu $}
  \put(26.3,31)  {\scriptsize $ \mu $}
  \put(43,10.8)  {\scriptsize $ \epsilon $}
  \put(43,20.1)  {\scriptsize $ \epsilon $}
  \put(58.3,0)   {\scriptsize $ \mu $}
  \put(58.1,31)  {\scriptsize $ \mu $}
  }
  \end{picture}
  $$
and the correlator of two spin fields on a torus can be expressed as
  $$
  \begin{picture}(393,135)
  \put(0,70)    {\includegraphics{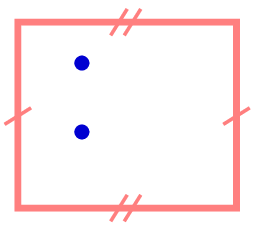}}
  \put(27.1,95.4){\scriptsize $\sigma$ }
  \put(27.1,115.4){\scriptsize $\sigma$ }
  \put(90,99)  {$ = $}
  \put(113,99) {$ \displaystyle \frac12 $}
  \put(129,70) {
  \put(0,0)    {\includegraphics{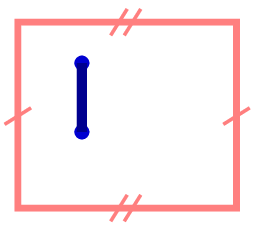}}
  \put(17.2,35){\scriptsize $\epsilon$ }
  \put(27.1,26){\scriptsize $\mu$ }
  \put(27.1,45.9){\scriptsize $\mu$ }
  }
  \put(216,99) {$ + $}
  \put(233,99) {$ \displaystyle \frac12 $}
  \put(251,70) {
  \put(0,0)    {\includegraphics{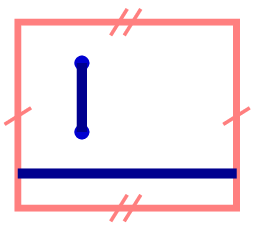}}
  \put(17.2,35){\scriptsize $\epsilon$ }
  \put(27.1,26){\scriptsize $\mu$ }
  \put(27.1,45.9){\scriptsize $\mu$ }
  \put(44,18.2){\scriptsize $\epsilon$ }
  }
  \put(130,29) {$ + $}
  \put(147,29) {$ \displaystyle \frac12 $}
  \put(165,0) {
  \put(0,0)    {\includegraphics{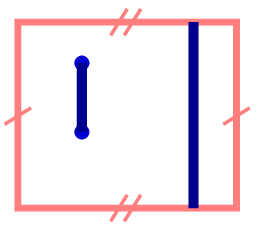}}
  \put(17.2,35){\scriptsize $\epsilon$ }
  \put(27.1,26){\scriptsize $\mu$ }
  \put(27.1,45.9){\scriptsize $\mu$ }
  \put(49.2,18){\scriptsize $\epsilon$ }
  }
  \put(252,29) {$ + $}
  \put(269,29) {$ \displaystyle \frac12 $}
  \put(287,0) {
  \put(0,0)    {\includegraphics{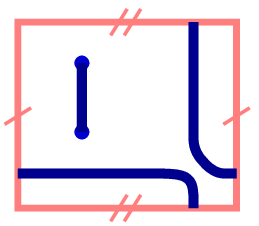}}
  \put(16.8,35){\scriptsize $\epsilon$ }
  \put(27.1,26){\scriptsize $\mu$ }
  \put(27.1,45.9){\scriptsize $\mu$ }
  \put(34.3,18.1){\scriptsize $\epsilon$ }
  \put(50,32){\scriptsize $\epsilon$ }
  }
  \end{picture}
  $$

T-duality of compactified free bosons can be understood as a special type of
order-disorder duality, see Section 5.4 of \cite{FGRS07}. Indeed, the chiral
data of two T-dual full conformal field theories can be described by the same
theory, namely the one describing a $\mathbb Z_2$-orbifold of the free boson theory. The 
line defect implementing the duality is then constructed
from twist fields of this orbifold theory.


\subsection{Relative field theories}

Before turning to defects in the specific class of three-dimensional Dijkgraaf-Witten theories,
we briefly explain two applications of defects. The first is in the context of {\em relative 
field theories}. By definition, a $(d{-}1)$-di\-men\-sional relative field theory is a field theory 
on the boundary of a $d$-di\-men\-sional field theory or, more generally, on a codimension-1 defect.
There are interesting cases in which the latter 
is a topological field theory; we mention two particular situations:
\\[3pt]
(1)\,
The case that the $d$-dimensional topological field theory is even an invertible theory.
This provides a geometric setup for describing anomalous theories; see \cite{FT14}.
\\[3pt]
(2)\,
The case $d \,{=}\, 3$, with the three-dimensional topological field theory being
of Reshetikhin-Turaev type and built from a modular tensor category that is 
the representation category of a vertex algebra. In this case the relative theory on a 
topological surface defect is a local rational conformal field theory
whose underlying chiral theory is the one built from the vertex algebra.
This provides a geometric interpretation \cite{KS11} of the TFT construction \cite{SFR06} 
of RCFT correlators.


\subsection{Quantum codes}

Another application of surface defects in three-di\-men\-sional topological field
theories are quantum codes. It is well known that the
space of qubits of the toric code on a surface $\Sigma$ can be described by
a topological field theory of Turaev-Viro type. Indeed, the vector space assigned by 
the topological field theory to the surface $\Sigma$ contains
the information on a vector space together with an action of mapping class groups 
which is used to construct quantum gates. Two problems are apparent: First, the
surfaces $\Sigma$ of concrete samples typically have a low genus, and thus, 
according to the Verlinde formula, the TFT state space space $\tft(\Sigma)$ has
low dimension. Moreover, the representation of the mapping class group might be too
small to allow for universal gates. To circumvent these problems the idea has been
put forward \cite{BW10,BQ12,BJQ13} to consider samples with a bilayer 
(or multi-layer) system and twist defects that create branch cuts. They effectively
lead to systems defined on surfaces of higher genus and thus to higher-di\-men\-sional
quantum codes with richer representations of mapping class groups.
Such systems have been analyzed mathematically \cite{FS14} by exploiting 
results on permutation equivariant categories.


\section{Defects and boundary conditions in three-dimensional topological field theories} 
\label{DWbound}

We now turn our attention to a particularly accessible subclass of three-di\-men\-sional
topological field theories, Dijkgraaf-Witten theories. These are topological field 
theories of Turaev-Viro type, and they can be constructed explicitly in a 
gau\-ge-the\-o\-retic setting with finite gauge group. We treat them as 
3-2-1-extended topological field theories.

\subsection{Constructing extended Dijkgraaf-Witten theories from $G$-bun\-dles}
\label{DWsect}

As a first input datum we select a finite group $G$. 
The space of field configurations on a closed oriented compact smooth
three-ma\-ni\-fold $M$ is then taken to be the groupoid $\Bun_G(M)$ of $G$-bundles on $M$. 
By considering formally a path integral with vanishing action, we obtain a partition function
that evaluates to
  $$
  \tft_G(M)=\int_{\Bun_G(M)}\!\! \mathrm D A\,\, \mathrm{e}^0 = \big| \Bun_G(M)\big| \,.
  $$
Here we use the fact that the groupoid $\Bun_G(M)$ is essentially finite, so that there is a 
well-defined counting measure, its groupoid cardinality. The groupoid cardinality of 
an (essentially) finite groupoid $\Gamma$ is
  $$
  \big| \Gamma \big| := \sum_{\gamma\in \pi_0(\Gamma)} \frac1{|\mathrm{Aut}_\Gamma(\gamma)|} \,,
  $$
where the summation is over the set $\pi_0(\Gamma)$ of isomorphism classes of objects of
$\Gamma$, each of which is counted with the inverse of the cardinality of its automorphism
group. 

Trivially, the number $\tft_G(M) \,{=}\, |\Bun_G(M)|$ defines an invariant of the 
three-manifold $M$.
It is non-trivial, though, that this invariant is local in the following sense. 
Let us cut the closed three-manifold into two pieces, consisting of
three-ma\-ni\-folds with boundaries that are closed oriented surfaces. Suppose that
the topological field theory can associate meaningful quantities to such manifolds
as well, which would allow us to reduce the problem of computing a three-manifold invariant
to computing invariants of simpler manifolds. To see what the topological field theory 
$\tft_G$ should associate to a closed oriented two-manifold $\Sigma$, we take a 
three-ma\-nifold $M$ with boundary
$\Sigma$. For any $G$-bundle $P$ on $\Sigma$, consider the groupoid $\Bun_G(M,P)$ of 
$G$-bundles on $M$ that restrict to the $G$-bundle $P$ on the boundary $\partial M$. Then
$\Bun_G(M,P)$ is an essentially finite groupoid: its groupoid cardinality provides a function
  $$
  \begin{array}{rll} 
  \Psi_M:\quad \Bun_G(\Sigma) &\!\! \longrightarrow \!\!& \mathbbm C\\[3pt]
  P &\!\! \longmapsto \!\!& |\Bun_G(M,P)|
  \end{array}
  $$
that depends only on the isomorphism class of the bundle $P$. To the surface $\Sigma$ itself, 
we should associate the recipient for all these functions: the vector space 
$\mathbbm C[\pi_0(\Bun_G(\Sigma))]$ of complex-valued functions on the set $\pi_0(\Bun_G(\Sigma))$ 
of isomorphism classes of $G$-bundles on $\Sigma$. We thus recover the well known 
feature that a three-dimensional topological field theory assigns vector spaces to 
surfaces. In the case of Dijkgraaf-Witten theories, these vector spaces are obtained by 
linearization of (isomorphism classes of objects of) categories of bundles.

Closed two-manifolds can, in turn, be decomposed, e.g.\ in a pair-of-pants decomposition, by
cutting them along circles. Hence let $S$ be a closed oriented one-manifold and ask
what an extended topological field theory should associate 
to $S$. Following the same type of analysis as above, we choose a surface
$\Sigma$ with boundary $S$. Fixing a $G$-bundle $P$ on the boundary $S$, we assign to it a vector
space,
  $$
  \Psi_\Sigma:\quad P \,\longmapsto\, \mathbbm C[\pi_0(\Bun_G(\Sigma,P))] \,.
  $$
The so obtained map $\Psi_\Sigma\colon \Bun_G(S)\,{\to}\, \Vect $ is a vector bundle on the space of
field configurations on $S$. Thus the extended
Dijkgraaf-Witten theory based on $G$ should associate to $S$ the $\mathbbm C$-li\-ne\-ar category 
of vector bundles over the space of field configurations. 
Denoting, for an essentially finite groupoid $\Gamma$, the functor
category from $\Gamma$ to a category $\mathcal C$ by $[\Gamma,\mathcal C]$,
this is the functor category
  $$
  \tft_G(S) = [\Bun_G(S),\Vect] \,.
  $$
We thus learn that a 3-2-1-extended topological field theory assigns to a one-ma\-ni\-fold 
a $\mathbbm C$-linear category. 
The category $\tft_G(S^1)$ can be computed explicitly: the category $\Bun_G(S^1)$
is equivalent to the action groupoid $G/\!/G$ for the adjoint action of $G$ on itself. Indeed,
after fixing a point $p \,{\in}\, S^1$, a $G$-bundle is described by its holonomy $g \,{\in}\, G$, 
on which gauge transformations act by the adjoint action. Thus $\tft_G(S^1) \,{\simeq}\,
[G/\!/G,\Vect]$, and this functor category is equivalent to the category of $G$-graded vector spaces
$V \,{=}\, \bigoplus_{g\in G} V_g$ with a $G$-action such that $g(V_h) \,{\subset}\, V_{ghg^{-1}}$. 
Put differently, we deal with $G$-equivariant vector bundles on the group $G$.

This category is actually nothing but the {\em Drinfeld center} of the fusion category
$G$-$\Vect$ of $G$-graded vector spaces. For later reference, we recall the definition of 
the Drinfeld center $\calz(\cala)$ of a monoidal category $\cala$: an object of $\calz(\cala)$
is a pair $(U,c_{U,-})$ consisting of an object $U$ of $\cala$ and a functorial family of
isomorphisms $c_{U,X}\colon U \,{\otimes}\, X \,{\to}\, X \,{\otimes}\, U$,
called a half-braiding, subject to some further coherence properties; a morphism from 
$(U,c_{U,-})$ to $(V,c_{V,-})$ is a morphism $f\colon U \,{\to}\, V$ in $\cala$ such that 
$(\mathrm{id}_- \,{\otimes}\, f) \,{\circ}\, c_{U,-} \,{=}\, c_{V,-} \,{\circ}\, (f \,{\otimes}\, 
\mathrm{id}_-)$. The half-braidings endow the Drinfeld center $\calz(\cala)$ with the structure 
of a {\em braided} monoidal category. Dropping the half-braiding gives a forgetful functor
  $$\begin{array}{rll}
  \calz(\cala) &\!\! \to \!\!& \cala\\[2pt]
  (U,c_{U-}) &\!\! \mapsto \!\!& U
  \end{array}$$
which has a natural monoidal structure. If a category $\calc$ is already braided, then the 
braiding provides a braided monoidal functor $\calc \,{\to}\, \calz(\calc)$. The opposite 
braiding endows the underlying category with another structure of a braided category, 
denoted by $\calc^{\mathrm{rev}}$, and similarly gives a braided monoidal functor 
$\calc^{\mathrm{rev}} \,{\to}\, \calz(\calc)$. These two functors combine into a braided 
monoidal functor 
  \begin{equation} 
  F:\quad \calc\boxtimes\calc^{\mathrm{rev}}\to \calz(\calc)\
  \label{cschweigert:cantriv}
  \end{equation}
from the Deligne product of $\calc$ and $\calc^{\mathrm{rev}}$ to the center.
A braided category is called \emph{non-degenerate} iff this functor $F$ is a braided 
equivalence. A non-degenerate ribbon category is called a \emph{modular} category.  

The preceding considerations for Dijkgraaf-Witten theories lead us naturally to
the general definition 
of 3-2-1-extended topological field theories as a symmetric monoidal 2-functor
  $$
  \tft:\quad \cobord_{3,2,1} \to 2\mbox-\Vect(\mathbbm C) \,.
  $$
This definition naturally generalizes Atiyah's classical definition of (non-extended)
topological field theory. 
In the rest of this subsection we elaborate on this 2-func\-tor. In the definition,
$2$-$\Vect(\mathbbm C)$ is the symmetric monoidal bicategory 
of finitely semisimple $\mathbbm C$-linear abelian categories, with the Deligne product 
as a mo\-no\-i\-dal structure. The bicategory $\cobord_{,3,2,1}$ is
an extended cobordism category whose objects are closed oriented one-manifolds $S$,
whose 1-morphisms $S \,{\to}\, S'$ are two-ma\-ni\-folds $\Sigma$ with boundary together
with a decomposition $\partial\Sigma \,{\xrightarrow\simeq}\, S'\,{\sqcup}\, {-}S$, and
whose 2-morphisms are three-manifolds with corners, up to diffeomorphism, again with an
appropriate decomposition of the boundary.

According to the definition, the 2-functor $\tft$ assigns to the oriented circle 
${\mathbb S}^1$ a finitely semisimple $\mathbbm C$-linear category 
${\mathcal C} \,{:=}\, \tft({\mathbb S}^1)$. The trinion (pair of pants), regarded as a 
cobordism ${\mathbb S}^1 \,{\sqcup}\, {\mathbb S}^1 \,{\to}\, {\mathbb S}^1$ gives a functor 
$\otimes\colon\, {\mathcal C}\,{\boxtimes}\, {\mathcal C} \,{\to}\, {\mathcal C}$. Finally, 
suitable three-manifolds with corners provide natural isomorphisms that 
endow this functor with an associativity constraint and the category ${\mathcal C}$ with 
the structure of a braiding. The constraints on these natural transformations, expressed
by pentagon and hexagon diagrams, can be deduced from homotopies between three-manifolds 
with corners. In this way the category ${\mathcal C} \,{=}\, \tft({\mathbb S}^1)$ is 
expected to be endowed with the structure of a modular tensor category
(see \cite{BDSV14} for some recent progress). The Reshetikhin-Turaev 
construction can be seen as a converse, constructing a 3-2-1-extended topological
field theory from a modular tensor category.


\subsection{Topological field theories with boundaries and defects}\label{sec:defbound}

Incorporating defects and boundaries into a topological field theory amounts to consider
an enlarged bicategory $\cobord^\partial_{3,2,1}$ of cobordisms. 
As a warm-up, we recall the situation for two-dimensional topological field theories
with boundaries, also known as open/closed topological field theory 
 \cite{La01,AN06,MS06,LP08}
Beyond disjoint unions of oriented circles, the category $\cobord^\partial_{2,1}$ has intervals
$\mathbb I_{a_1,a_2}$, with boundary conditions $a_1,a_2$ attached to the end points,
as additional objects. Morphisms are now surfaces with boundaries consisting of segments 
which are are either ``cut-and-paste  boundaries'' implementing locality or true physical boundaries.

A two-dimensional open/closed topological field theory is a symmetric monoidal functor 
$\tft\colon \cobord^\partial_{2,1}\to\Vect$. It assigns a vector space 
  $$
  A_a := \tft(\mathbb I_{aa})
  $$
to an interval with boundary condition $a$ at either end point. 
For any boundary condition $a$, a disk with three marked intervals on the boundary provides 
a cobordism $\mathbb I_{aa} \,{\sqcup}\, \mathbb I_{aa} \,{\to}\, \mathbb I_{aa}$; applying 
the functor $\tft$ thus yields a linear map $A_a \,{\otimes}\, A_a \,{\to}\, A_a$. 
Further analysis \cite{MS06,LP08} shows that this map furnishes a (not necessarily commutative)
associative product on $A_a$, and that $A_a$ is even endowed
with the structure of a symmetric Frobenius algebra. A similar argument shows that for any 
other boundary condition $b$, the cobordism $\mathbb I_{aa}\sqcup \mathbb I_{ab} \,{\to}\, \mathbb I_{ab}$ 
endows the vector space $M_b \,{:=}\, \tft(\mathbb I_{ab})$ with a linear map 
$A_a \,{\otimes}\, M_b \,{\to}\, M_b$ and thereby acquires the structure of an $A_a$-module.

In this way one arrives at the following structure: boundary conditions form a $\mathbbm C$-linear 
category that is equivalent to the category $A_a\text{-mod}$ of $A_a$-modules for any given 
boundary condition $a$. To the interval $\mathbb I_{b_1,b_2}$ the topological field theory assigns 
  \begin{equation}
  \tft(\mathbb I_{b_1,b_2})= \mathrm{Hom}_{A_a\text{-mod}}(b_1,b_2) \,,
  \label{cschweigert:interval}
  \end{equation}
the vector space of morphisms of $A_a$-modules. In particular, for any boundary condition
$b$, we find $\tft(\mathbb I_{b,b}) \,{=}\, \mathrm{End}_{A_a\text{-mod}}(b)$, which is an algebra that 
is Morita equivalent to the algebra $A_a$. If $A_a$ is semisimple, then the commutative Frobenius algebra 
$C \,{:=}\, \tft(S^1)$ that is assigned to the circle turns out to be the center of the algebra $A_a$:
  $$
  C= \tft(S^1)= Z(\tft(\mathbb I_{aa}) )= Z(A_a) \,.
  $$

For three-dimensional topological field theories, we will find a higher-categorical analogue 
of these structures. In a complete treatment one should start by giving a precise definition 
for the extended category $\cobord^\partial_{3,2,1}$ of cobordisms;
several definitions have been proposed for this enlarged category, see e.g.\ \cite[Sect.\ 4.3]{Lu09}.
Here we refrain from presenting a formal definition, and instead concentrate on some of the physical 
requirements that such a category should satisfy. Again we start from oriented three-manifolds, but this 
time we also allow them to have  oriented
boundaries that are actual physical boundaries to which boundary 
conditions need to be assigned, rather than the cut-boundaries that were already present in 
Section \ref{DWsect}, which are instead a device for incorporating some aspects of locality into 
the theory. We also admit codimension-1 submanifolds, called surface defects. Such a defect separates
two three-dimensional regions each of which supports a topological field theory, and the 
theories associated to the two regions may be different. Labels must be assigned to
such defects, too. In this way, we get in particular invariants of three-manifold with embedded
surfaces. In the present setting of 3-2-1-extended topological field theories, 
it is natural to go one step further and allow for
codimension-2 defects as well. In particular, there will be generalized Wilson lines
separating two surface defects from each other, and likewise boundary Wilson lines that are 
confined to a boundary component.

In a next step one imposes locality by cutting such a three-manifold to produce three-ma\-nifolds 
with cut-boundaries. It is natural to impose the condition that the cut-surface intersects all
structure in the three-manifold, e.g.\ surface defects and boundaries, transversally. Boundary manifolds
now carry a decoration consisting of lines labeled by surface defects and boundaries labeled
by boundary conditions. Imposing further locality, as in Section \ref{DWsect}, requires to
cut such decorated surfaces transversally. This yields new types of labeled one-manifolds, which are to
be taken as the objects in the bicategory $\cobord^\partial_{3,2,1}$. Let us describe these objects
in detail: There are oriented intervals and circles, with finitely many marked points in the interior 
that divide them into subintervals. Each subinterval is to be marked with a topological field theory.
Points adjacent to two subintervals are to be marked by a surface defect, boundary points by
a boundary condition. The possible boundary conditions and types of surface defects have to be
determined by a field-theoretic analysis. 
For the case of Dijkgraaf-Witten theories we present this analysis in the next subsection.


\subsection{A model independent analysis of defects and 
boundaries}\label{sec:tftb}

We now turn to a general analysis of boundary conditions and defects in three-dimensional
topological field theories of Reshetikhin-Turaev type. 
The labels for the bulk Wilson lines in such a theory are supplied by some modular tensor category $\calc$.
The boundary conditions of the theory
form a bicategory: its objects are boundary conditions, its 1-morphisms are labels for
boundary Wilson lines, and its 2-morphisms label point-like insertions on boundary Wilson lines.
Our goal is to describe this bicategory in terms related to the modular tensor category $\calc$.

We want the boundary Wilson lines for a given boundary condition to be topological as well. 
As a consequence there is a fusion of boundary Wilson lines and we have invariance under two-dimensional
isotopies. This leads to the requirement that for any boundary condition $a$ the 
boundary Wilson lines and their insertions form a $\mathbbm C$-linear, finitely semisimple, rigid monoidal
category $\calw_a$. The category $\calw_a$ is thus a fusion category; since boundary Wilson lines are
confined to the two-dimensional boundary, 
$\calw_a$ is, however, not endowed with the structure of a braiding.

To derive constraints on the category $\calw_a$, we postulate that there should exist a process 
of moving a bulk Wilson line to the boundary, and that this should be as smooth as possible.
(In other words, we exclude possible boundary conditions for which such a smooth process does 
not exist.) This provides a functor
$F_a\colon \calc \,{\to}\, \calw_a$. Next we impose the condition that the following two processes 
are equivalent: fusing two bulk Wilson lines and then moving them into the boundary, and moving 
the two bulk Wilson lines individually to the boundary and afterwards fusing them in the boundary; 
schematically,
  $$ 
  \begin{picture}(350,88)
  \put(0,3)     {
  \put(0,0)     {\includegraphics{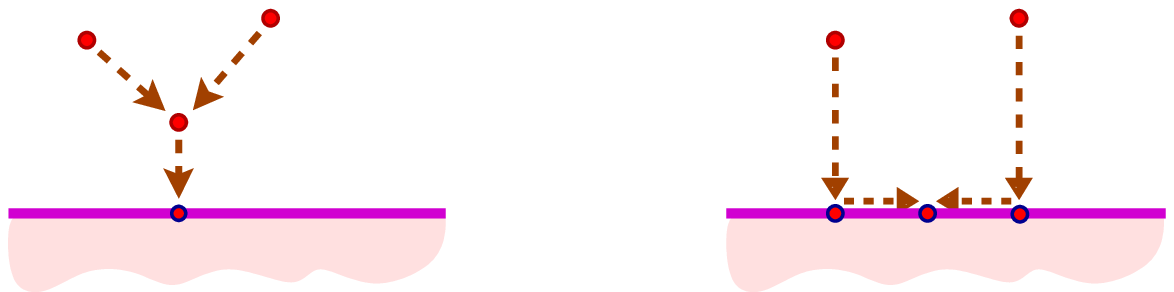}}
  \put(15,71)     {\scriptsize $ U $}
  \put(56.5,46.3) {\scriptsize $ U{\otimes}V $}
  \put(54,26)     {\begin{turn}{15} \scriptsize $ F_a(U{\otimes}V) $\end{turn}}
  \put(82.7,78)   {\scriptsize $ V $}
  \put(216.5,0){
  \put(-1,27.7)   {\scriptsize $ F_a(U) $}
  \put(15,71)     {\scriptsize $ U $}
  \put(48.3,26.3) {\begin{turn}{80} \scriptsize $ F^{}_{\!a}(U)\,{\otimes}\, F^{}_{\!a}(V) $\end{turn}}
  \put(82.7,78)   {\scriptsize $ V $}
  \put(82,27.7)   {\scriptsize $ F_a(V) $}
  } }
  \end{picture}
  $$
In formulas, what we postulate is the existence of coherent isomorphisms 
  $$
  F_a(U{\otimes_\calc} V) \cong F_a(U) \,{\otimes_{\calw_a}}\, F_a(V)
  $$
for all $U,V \,{\in}\,\calc$. This means that the functor $F_a$ has the structure of a 
\emph{monoidal} func\-tor.

Similarly, the following two processes should be equivalent:
  $$ 
  \begin{picture}(350,87)
  \put(0,3)     {
  \put(0,0)     {\includegraphics{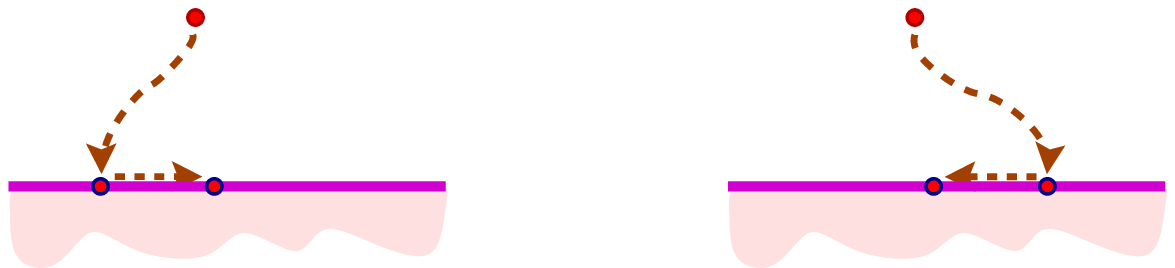}}
  \put(2.9,34.1)  {\begin{turn}{-20}$\scriptstyle F_a(U)$\end{turn}}
  \put(61.9,70.8) {\scriptsize $ U $}
  \put(59,28.2)   {\scriptsize $ X $}
  \put(254.5,71)  {\scriptsize $ U $}
  \put(266.1,28.2){\scriptsize $ X $}
  \put(304.6,26.2){\begin{turn}{20}$\scriptstyle F_a(U)$\end{turn}}
  }
  \end{picture}
  $$
Here a bulk Wilson line labeled by an object $U$ of $\calc$ is moved to either side of a boundary Wilson line
labeled by an object $X$ of $\calw_a$. Accordingly we postulate the existence of coherent isomorphisms
  $$
  F_a(U) \,{\otimes_{\calw_a}}\, X \cong X \,{\otimes_{\calw_a}}\, F_a(U) \,.
  $$
This constitutes an additional structure on a monoidal functor from a braided category
to a fusion category, called the structure of a \emph{central}
functor. It can be alternatively described as the choice of a lift $\widetilde F_a$ of $F_a$
to the center of $\calw_a$, i.e.\ as the existence of a braided monoidal functor $\widetilde F$
such that the diagram
  $$\xymatrix{
  & \calz(\calw_a) \ar[d]\\
  \calc \ar^{F_a}[r] \ar@{-->}^{\widetilde F_a}[ur]& \calw_a
  }$$
commutes, with the vertical arrow the forgetful functor described in Section \ref{DWsect}.
It is now  an evident naturality requirement to ask that $\widetilde F$ is an \emph{equivalence}
of braided categories: the only reason for the existence of a consistent prescription for 
moving a specific boundary Wilson line past any arbitrary boundary Wilson line
should be that this boundary Wilson line actually comes from
a Wilson line in the bulk via the process described by the functor $F_a$.

We thus learn in particular that in order to admit boundary conditions of the type just introduced,
a modular tensor category $\calc$ must be a Drinfeld center. This constitutes a new
phenomenon in three dimensions, for which there is no
counterpart in two dimensions: while any commutative algebra 
(arising, as explained in Section \ref{sec:defbound}, as the value of a two-dimensional topological field
theory on a circle) is the center of some algebra (namely, of itself), it is {\em not\/}
true that any braided monoidal fusion category (arising as the value of a
three-dimensional topological field theory on a circle) is the Drinfeld center of a fusion category.

In this context, the following notions are relevant \cite{DMNO13,DNO13}: Two modular
tensor categories $\calc_1$ and $\calc_2$ are said to be {\em Witt equivalent\/} iff there
exist fusion categories $\cala_1$ and $\cala_2$ and a braided equivalence
  $$
  \calc_1\boxtimes \calz(\cala_1) \xrightarrow{~~\simeq~~} \calc_2\boxtimes\calz(\cala_2) \,.
  $$
The Deligne product induces on the set of equivalence
classes a natural group structure; this group is called the Witt group 
of modular tensor categories. It contains, as a subgroup, the Witt group of finite abelian
groups $A$ that are endowed with a quadratic form $q\colon A \,{\to}\, {\mathbb C}^\times$.
(Such a form determines a modular tensor category, a fact that is relevant 
for abelian Chern-Simons theories.) The obstruction to the existence of boundary conditions
for a three-dimensional topological field theory based on a modular tensor category
$\calc$ can thus be restated as the condition that the class of $\calc$ in the Witt group must
vanish, $[\calc] \,{=}\, 0$. 

Suppose now we have already found one boundary condition $a$, i.e.\ one Witt-tri\-vi\-alization
      $\widetilde F_a\colon \calc \,{\xrightarrow{\,\simeq\,}}\, \calz(\calw_a)$. If $b$ is any other boundary 
condition, denote by $\calw_{a,b}$ the category of boundary Wilson lines separating a piece of 
boundary in boundary condition $a$ from a piece in boundary condition $b$. Fusion of Wilson lines
inside the piece in boundary condition $a$ with a Wilson line separating $a$ and $b$ then
yields a functor
  $$
  \calw_a \,{\times}\, \calw_{a,b} \longrightarrow \calw_{a,b} \,.
  $$
Hereby the category $\calw_{a,b}$ acquires the structure of a module category over the monoidal category
$\calw_a$. This amounts to a categorification of the situation in two-dimensional topological
field theory analyzed in Section \ref{sec:defbound}. The bicategory of module categories
over $\calw_a$ is therefore a natural candidate for the bicategory of boundary conditions.

In much the same way as a left module $M$ over a ring $R$ is a right module over the
ring $\mathrm{End}_R(M)$, the $\calw_a$-module category $\calw_{a,b}$ is also a right
module category over the monoidal category ${\mathcal E\!nd}_{\calw_a}(\calw_{a,b})$
of module endofunctors of $\calw_{a,b}$. But at the same
time it is also a right module category over the fusion category $\calw_b$ of Wilson lines inside 
the boundary with boundary condition $b$. Naturality thus suggests
an equivalence $\calw_b\,{\simeq}\, {\mathcal E\!nd}_{\calw_a}(\calw_{a,b})$. This can only hold true 
if the fusion category ${\mathcal End}_{\calw_a}(\calw_{a,b})$ is a Witt trivialization of $\calc$ as
well. Now indeed it is known \cite{Sc01} that for any module category there is a
canonical braided equivalence 
  $$
  \calz(\calw_a) \simeq \calz({\mathcal E\!nd}_{\calw_a}(\calw_{a,b})) \,,
  $$
so that we do have a canonical Witt trivialization for ${\mathcal E\!nd}_{\calw_a}(\calw_{a,b})$ as well.

More generally, given two boundary conditions described by $\calw_a$-module category $\calm_b$ and
$\calm_c$, the category of boundary Wilson lines is given by the functor category 
${\mathcal F\!un}_{\calw_a}(\calm_b,\calm_c)$ of module functors. This can be regarded as a categorification
of Equation (\ref{cschweigert:interval}). At this point a word of warning is in order: via pullback along
the forgetful functor
  $$
  \calc \xrightarrow{~\simeq~} \calz(\calw_a) \xrightarrow{~\phantom{\simeq}~}  \calw_a \,,
  $$
any $\calw_a$-module category has a natural structure of a $\calc$-module category; 
however, not every $\calc$-module category describes a boundary condition for the 
Reshetikhin-Turaev theory based on $\calc$. As an illustration consider Kitaev's toric
code, which is the Dijkgraaf-Witten theory based on $\mathbb{Z}_2$. In that case there are 
$6$ inequivalent 
    indecomposable
$\calc$-module categories over the Drinfeld center 
$\calc\,{=}\,\calz(\Vect(\mathbb Z_2))$ of the category $\Vect(\mathbb Z_2)$ of 
$\mathbb Z_2$-graded complex vector spaces, but only 2 
   indecomposable
module categories over the fusion category $\Vect(\mathbb Z_2)$. These two 
$\Vect(\mathbb Z_2)$-module categories correspond to the two known \cite{brki} 
   elementary
boundary conditions for the toric code.

The results obtained above are easily extended to surface defects that separate 
three-dimensional topological field theories of Reshetikhin-Turaev type based on two 
modular tensor categories $\calc_1$ and $\calc_2$. Again, one starts with a fusion category 
$\calw_d$ of defect Wilson lines that are confined to the 
defect surface. We now get two monoidal functors
  $$
  \calc_1\xrightarrow{~~} \calw_d\qquad\text{and}\qquad 
  \calc_2^{\mathrm{rev}} \xrightarrow{~~} \calw_d
  $$
which combine into a single functor
  $$
  F_d:\quad \calc_1 \,{\boxtimes}\, \calc_2^{\mathrm{rev}} \xrightarrow{~~~} \calw_d \,.
  $$
Again this functor $F_d$ has the structure of a central functor, yielding a functor
  $$
  \widetilde F_d:\quad \calc_1 \,{\boxtimes}\, \calc_2^{\mathrm{rev}} \xrightarrow{~~\simeq~~} \calz(\calw_d)
  $$
that is again argued to be, by naturality, a braided equivalence. This leads to the following
obstruction: surface defects of the type studied here only exist if the classes of the
two modular tensor categories in the Witt group are equal, $[\calc_1] \,{=}\, [\calc_2]$.

Our discussion is, deliberately, not at the highest possible conceptual level.
In analogy with the fact that
Morita equivalent algebras can be naturally discussed in a bicategorical setting with
bimodules as 1-morphisms, in the present situation there is an underlying 3-categorical structure. 
Also, we did not
discuss fusion of topological surface defects. Indeed, surface defects form a monoidal bicategory.
The monoidal unit is called the {\em transparent surface defect}, which separates a modular
tensor category $\calc$ from itself. The underlying Witt trivialization is based
on $\calc$, seen as a fusion category, and the braided equivalence
  $$
  \calc\boxtimes \calc^{\mathrm{rev}} \xrightarrow{~~\simeq~~} \calz(\calc)
  $$
which follows by Equation (\ref{cschweigert:cantriv}) 
from the fact that $\calc$ is modular. Ordinary Wilson lines are then
recovered as defect Wilson lines separating the transparent defect from itself.


\subsection{Defects and boundaries in Dijkgraaf-Witten theories}\label{sec:DW}

We now describe an explicit gauge-theoretic construction \cite{FSV14} of boundary conditions and
surface defects for Dijkgraaf-Witten theories. A key idea is to keep the two-step
procedure outlined in Section \ref{DWsect}: first we assign to manifolds of various dimension
groupoids of (ge\-nera\-lized) bundles, and then we linearize those categories.

Our ansatz is inspired by the notion of a {\em relative bundle}: Given a morphism
$j\colon Y \,{\to}\, X$ of smooth manifolds and a group homomorphism $\iota\colon H \,{\to}\, G$, 
the category of relative bundles has as objects triples consisting of a $G$-bundle $P_G \,{\to}\, X$,
an $H$-bundle $P_H \,{\to}\, Y$, and an isomorphism 
  $$
  \alpha:\quad \mathrm{Ind}_H^G (P_H) \,\xrightarrow{\,\,\cong\,\,}\, j^* P_G
  $$
of $G$-bundles on $Y$. A morphism is a pair consisting of a morphism 
$P_G \,{\xrightarrow{\,\varphi_G^{}\,}}\, P_G'$ of $G$-bundles on $X$ 
and a morphism $P_H \,{\xrightarrow{\,\varphi_H^{}\,}}\, P_H'$ on $Y$
of $H$-bundles, obeying the obvious compatibility constraint 
  $$\xymatrix{
    \mathrm{Ind}_H^G (P_H^{})\ar^{~~~\alpha}[r]\ar_{\mathrm{Ind}_H^G(\varphi_H^{})} [d]
    & j^*P_G^{}\ar^{j^*\!\varphi_G^{}}[d]\\
  \mathrm{Ind}_H^G (P_H')\ar^{~~~\alpha'}[r] &j^*P_G'
  }$$
with the isomorphisms $\alpha$ and $\alpha'$.

In our construction, relative bundles are the appropriate tool for describing boundaries. In the 
case of a surface defect $\Sigma \,{\hookrightarrow}\, M$ the situation is somewhat more subtle.
Suppose that $\Sigma$ divides the three-manifold $M$ into two disjoint connected components $M_\pm$. 
We consider the situation that Dijk\-graaf-Witten theories for two, possibly different, finite 
groups $G_\pm$ are associated to the components $M_+$ and $M_-$. Denote by 
$\overline M_\pm \,{:=}\, M_\pm \,{\cup}\, \Sigma$ the closure of $M_\pm$ in $M$, and by
$j_\pm\colon \Sigma \,{\hookrightarrow}\, \overline M_\pm$ the corresponding embeddings.
We then consider, for a given group homomorphism $H\to G_+ \,{\times}\, G_-$ as field 
configurations the groupoid whose objects consist of two $G_\pm$-bundles $P_\pm \,{\to}\, M_\pm$, 
an $H$-bundle $P_H \,{\to}\, \Sigma$ and an isomorphism 
  $$
  \beta :\quad \mathrm{Ind}_H^G(P_H) \,\xrightarrow{\,\,\cong\,\,}\,  j_+^*P_+^{}\times j_-^*P_-^{}
  $$ 
of $G_+{\times} G_-$-bundles on $\Sigma$.

We have to add yet one further datum to Dijkgraaf-Witten theories. Namely, typically, the 
linearization of groupoids is twisted by a 2-cocycle on the relevant groupoid of 
generalized bundles. This requires the choice of additional data on the finite groups involved. 
We first discuss the situation without boundaries and without surface defects, for a
Dijkgraaf-Witten theory based on a finite group $G$. Then
the additional datum is a three-cocycle $\omega \,{\in}\, Z^3(G,\mathbbm C^\times)$ on the group. 
A more geometrically inclined reader is invited to think about this cocycle as a three-cocycle
on the stack $\Bun_G$ of $G$-bundles, and thus as a (Chern-Simons) 2-gerbe on $\Bun_G$.
Such a three-cocycle leads to a holonomy on closed three-manifolds and thus furnishes a 
topological Lagrangian, yielding a three-dimensional topological field theory $\tft_{G,\omega}$.
To determine the category $\tft_{G,\omega}({\mathbb S}^1)$, we need a 2-cocycle $\tau(\omega)$
on the groupoid $\Bun_G(\mathbb{S}^1) \,{\simeq}\, G{/}\!{/} G$, the action groupoid for the 
adjoint action of $G$ on itself. The transgressed 2-cocycle 
$\tau(\omega)$ can be obtained \cite{Wi08} from the group cocycle $\omega$ by 
evaluating $\omega$ on a suitable triangulated three-ma\-ni\-fold. This indeed produces
the well-known 2-cocycle \cite{DPR90} for Dijkgraaf-Witten theories.

We now illustrate the generalization of this prescription to boundaries and defects in an 
example. Let $\mathbb I$ be an interval with a marked point in its interior, corresponding to
a surface defect. To each of the two subintervals ${\mathbb I}_{1,2}$ we a assign 
a finite group $G_{1,2}$ and three-cocycles 
$\omega_{1,2} \,{\in}\, Z^3(G_{1,2},\mathbbm C^\times)$, of which we think of as 
topological bulk Lagrangians. To the end point adjacent to the first interval,
we assign a group homomorphism $\iota_1\colon\, H_1 \,{\to}\, G_1$ and a boundary Lagrangian 
as follows. We think about the three-cocycle $\omega_1$ on $G_1$ as a Chern-Simons 2-gerbe on
$\Bun_{G_1}$. The group homomorphism $\iota_1$ induces a morphism
$\Bun_{H_1} \,{\to}\, \Bun_{G_1}$ of 2-ger\-bes. A similar situation, one categorical dimension lower,
is familiar from the study of D-branes, see e.g.\ \cite{FNSW10}: for D-branes, one has a gerbe on
the bulk manifold and a morphism of gerbes on the boundary, from the trivial gerbe 
to the restriction of the bulk gerbe. This is an instance of the general principle that
in gauge theories, holonomies of manifolds with boundary, and thus topological actions
on such manifolds, can only be defined if a trivialization on the boundary is chosen.

In the situation at hand, what is to be trivialized is the restriction of the Chern-Simons
2-gerbe to $\Bun_{H_1}$. Thus in terms of group cochains, we are looking for a 
2-cochain $\theta_1 \,{\in}\, C^2(H_1,\mathbbm C^\times)$ that represents a morphism
from the trivial 2-gerbe on $\Bun_H$ to the 2-gerbe described by the restriction of the
three-cocycle $\omega_1$ on $G_1$ to $\Bun_{H_1}$. Accordingly we
impose the condition ${\mathrm d}\theta_1 \,{=}\, \iota_1^*(\omega_1)$. We think about $\theta_1$
as a topological boundary Lagrangian. The situation at the other end point is, in complete
analogy, described by a group homomorphism $\iota_2\colon H_2 \,{\to}\, G_2$ and a 2-cochain
$\theta_2 \,{\in}\, C^2(H_2,\mathbbm C^\times)$ such that 
${\mathrm d}\theta_2 \,{=}\, \iota_2^*(\omega_2)$.

In a similar vein, the situation for the surface defect is a higher-categorical analogue 
of bibranes \cite{FSW08}, which describe the target space physics of topological defects
in two-dimensional conformal field theory: 
We now have a group homomorphism $\iota_{12}\colon H_{12} \,{\to}\, G_1\,{\times}\, G_2$ 
and a 2-cochain $\theta_{12} \,{\in}\, C^2(H_{12},\mathbbm C^\times)$ satisfying 
${\mathrm d}\theta_{12} \,{=}\, \tilde\iota_{2}^*(\omega_2) \,{\cdot}\, \tilde\iota_{1}^*(\omega_1)^{-1}$, 
with the group homomorphisms $\tilde\iota_i\colon H_{12} \,{\xrightarrow{\,\,\iota_{12}\,}}\,
G_1 \,{\times}\, G_2 \,{\xrightarrow{\,\,p_i\,}}\, G_i$ obtained from $\iota_{12}$ and the
canonical projections.
Indecomposable boundary conditions and defects correspond to group homomorphisms 
$\iota_{1,2}$, respectively $\iota_{12}$, that are injective, i.e.\ are subgroup embeddings.

In \cite{FSV14}, an explicit prescription was given  for transgressing these cochains to
a 2-cocycle on the relevant category of (generalized) relative bundles. This prescription
is a generalization of the one given in \cite{Wi08} for Dijkgraaf-Witten theories without 
boundaries or defects. It makes the categories associated to one-manifolds with
defects and boundaries quite explicitly computable.

On the other hand, there are also general results for boundary conditions and defects in
three-dimensional topological field theories of Reshetikhin-Turaev type. Suppose the
theory is based on a modular tensor category $\mathcal C$. The model independent analysis 
of Section \ref{sec:tftb} reveals that topological boundary conditions only exist
if the category $\mathcal C$ is braided equivalent to the Drinfeld center of a fusion
category $\mathcal A$. For Dijkgraaf-Witten theories based on a finite group $G$ and
a three-cocycle $\omega \,{\in}\, Z^3(G,\mathbbm C^\times)$, this condition is satisfied
automatically: the category $\mathcal C \,{=}\,\, \mathcal C_{G,\omega}$ is the 
Drinfeld center of the category $\Vect(G)^\omega$ of $G$-graded vector spaces 
with associator twisted by the three-cocycle $\omega$.

Recall from the general analysis in Section  \ref{sec:tftb}
that boundary conditions are in bijection
with module categories over the fusion category $\mathcal A$. For the fusion category
$\Vect(G)^\omega$ relevant for Dijkgraaf-Witten theories, indecomposable module categories are,
in turn, known \cite{Os03} to be in bijection with subgroups 
$\iota\colon H \,{\hookrightarrow}\, G$ and 2-cochains $\theta \,{\in}\, C^2(H,\mathbbm C^\times)$ 
such that ${\mathrm d}\theta \,{=}\, \iota^*\omega$. This matches exactly the gauge-theoretic
description of types of boundary conditions. According to the general results \cite{FSV13},
the topological field theory should assign to an interval with Dijkgraaf-Witten theory
based on $(G,\omega)$ and boundaries with boundary conditions $(H_1,\theta_1)$ and
$(H_2,\theta_2)$ the category 
  $$
  \mathrm{Fun}_{{\Vect(G)}^\omega}({\mathcal M}(H_1,\theta_1), {\mathcal M}(H_2,\theta_2))
  $$
of module functors from the indecomposable module category ${\mathcal M}(H_1,\theta_1)$ 
over the fusion category $\Vect(G)^\omega_{}$ to the indecomposable module category 
${\mathcal M}(H_2,\theta_2)$. A non-trivial calculation \cite{FSV14} shows that 
this category coincides, as a finitely semisimple abelian category, with
the category obtained by linearizing categories of generalized relative bundles.

A similar calculation can be performed for the circle with surface defects as point-like
insertions. This requires a higher-categorical version of the notion of a trace, namely 
a trace 2-functor on bimodule categories with values in $\mathbbm C$-linear categories. 
Such a coherent trace can indeed be constructed \cite[Sect.\,3]{fss} and gives rise
to results that are in accordance with
gauge theoretic calculations in Dijkgraaf-Witten theories \cite[Sect.\,5]{fss}.


\subsection{Symmetries from invertible topological surface defects}\label{sub:syminv}

We now turn to the relation between symmetries of three-dimensional topological field
theories of Turaev-Viro type and invertible surface defects. For such a theory, the modular
tensor category $\mathcal C$ of bulk Wilson lines is the Drinfeld center of a fusion category
$\mathcal A$, i.e.\ ${\mathcal C} \,{=}\, {\mathcal Z}({\mathcal A})$. 

According the paradigm discussed in Section 1, symmetries should correspond to invertible topological
surface defects, and their action should be described by wrapping those defects. By the
analysis presented in the preceding subsection, invertible surface defects are described
by invertible bimodule categories over the fusion category $\mathcal A$. These form a
monoidal bicategory; restricting to only invertible 1-morphisms and 2-morphisms in this bicategory, 
one obtains a categorical 2-group, the Brauer-Picard 2-group \cite{ENOM10}.
Its isomorphism classes form a finite group $\BrPic(\mathcal A)$; in the sequel
work with this group, rather than the underlying 2-group.

Applying an extended topological field theory to a two-manifold with boundary  
yields a functor. Take this two-manifold to be a
cylinder with a surface defect of type $D$ wrapping the non-contractible cycle:
  $$
  \begin{picture}(225,70)
  \put(20,1)  {\includegraphics{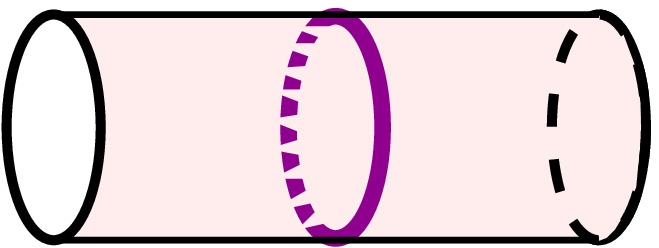}}
  \put(133.6,32) {\boldmath{$D$}}
  \end{picture}
  $$

This yields an endofunctor $F_D\colon\, {\mathcal C} \,{\to}\, {\mathcal C}$. A detailed
analysis \cite{FPSV14} shows
that if the defect $D$ is invertible, then the functor $F_D$ has the structure of a braided monoidal
autoequivalence. In the case of Dijkgraaf-Witten theories, the modular tensor
category of bulk Wilson lines is ${\mathcal C} \,{=}\, {\mathcal Z}(\Vect(G))$, 
and an invertible (and thus indecomposable) defect is given by a subgroup 
$\iota\colon H \,{\hookrightarrow}\, G\,{\times}\, G$ and a 2-cochain 
$\theta \,{\in}\, C^2(H,\mathbbm C^\times)$. The functor $F_D \,{\equiv}\, F_{(H,\theta)}$
can then be explicitly obtained by linearizing the span of action groupoids 
  $$ \xymatrix{
  &H/\!/H\ar^{\hat\pi_2}[dr]\ar_{\hat \pi_1}[dl]&\\ 
  G/\!/G && G/\!/G
  }$$
for adjoint actions which describe the relevant categories of 
(relative) bundles. Here $\hat \pi_i$ is the functor on groupoids that is induced from the 
group homomorphism $H\,{\xrightarrow{\,\iota\,}}\, G \,{\times}\, G\,{\xrightarrow{\,p_i\,}}\, G$.
One can then compute a pull-push functor on the linearizations to arrive at the functor
  $$ 
  F_{(H,\theta)}:\quad {\mathcal Z}(\Vect(G))=[G/\!/G,\Vect]
  \longrightarrow [G/\!/G,\Vect] \,,
  $$
which comes with a natural monoidal structure and which is braided. This construction
provides an explicit group homomorphism
  $$
  \BrPic(\mathcal A) \longrightarrow \mathrm{brdEq} ({\mathcal Z}(\Vect(G))) \,
  $$
from the Brauer-Picard group to the group $\mathrm{brdEq} ({\mathcal Z}(\Vect(G)))$ 
of braided autoequivalences of the Drinfeld center of $\Vect(G)$. A detailed comparison 
\cite{FPSV14} shows that this group homomorphism is exactly the one considered in 
\cite{ENOM10}; this embeds the construction of that paper naturally
into the framework of topological field theories with defects. It is also shown in \cite{ENOM10}
that the group homomorphism is actually an \emph{iso}morphism. According to the considerations
above, this representation theoretic result
can be reinterpreted as the statement that symmetries of topological field theories of
Turaev-Viro type can already be detected from their action on the category of bulk Wilson lines.


\subsection{Symmetries for abelian Dijkgraaf-Witten theories}

We now restrict our attention to the class of Dijkgraaf-Witten theories with vanishing 
three-cocycle and based on a finite abelian group $A$. In this case the Brauer-Picard group 
is explicitly known \cite{ENOM10}:
  $$
  \BrPic(\Vect(A)) \cong \mathrm{O}_q(A\oplus A^*)\,,
  $$
where $A^*$ is the character group of $A$, $q\colon A\,{\oplus}\, A^*\,{\to}\,\mathbbm C^\times$ 
is the natural quadratic form that is determined by $q(g,\chi) \,{=}\, \chi(g)$ for $g\,{\in}\,A$
and $\chi \,{\in}\, A^*$, and $\mathrm{O}_q(A{\oplus} A^*)$ is the subgroup of
those automorphisms of the finite group $A\,{\oplus}\, A^*$ that preserve this quadratic form $q$.

There are three obvious types of symmetries the Dijkgraaf-Witten theory should possess:
\\[4pt]
(1)\, \emph{Symmetries of} $\Bun_A$.
\\[2pt]
As explained in Section \ref{DWbound}, we can think about the stack $\Bun_A$ as a target space
for Dijkgraaf-Witten theories. Accordingly, symmetries of $\Bun_A$, i.e.\ group automorphisms of
the abelian group $A$, can be expected to be kinematical symmetries of the Dijkgraaf-Witten theory.
\\
Indeed, for $\varphi \,{\in}\, \mathrm{Aut}(A)$ the graph 
{\sl graph}$(\varphi)\,{\subset}\, A\,{\oplus}\, A$ is a subgroup; together with the trivial 
2-cochain $\theta \,{=}\, 1\,{\in}\, C^2(${\sl graph}$(\varphi),\mathbbm C^\times)$
it describes an invertible surface defect. Following the prescription given in Section
\ref{sub:syminv}, we compute the corresponding braided autoequivalence to be the element
  $$
  \varphi\oplus (\varphi^*)^{-1}:\quad A\oplus A^* \to A\oplus A^*
  $$
of $\mathrm{O}_q(A{\oplus} A^*)$.
\\[4pt]
(2)\, \emph{Automorphisms of the Chern-Si\-mons $2$-gerbe on} $\Bun_A$.
\\[2pt]
Recall that the three-cocycle $\omega\,{\in}\, C^3(A,\mathbbm C^\times)$ has the 
interpretation of a Chern-Simons 2-gerbe on $\Bun_A$. In the case at hand this 2-gerbe 
is trivial. Nevertheless its automorphisms are not trivial;
rather, they are given by 1-gerbes on $\Bun_A$.
Their isomorphism classs are given by the cohomology group $H^2(A,\mathbbm C^\times)$.
By transgression \cite{Wi08}, any cohomology class gives an alternating bihomomorphism 
$\beta\colon A \,{\times}\, A \,{\to}\, \mathbbm C^\times$, which in physical terms 
can be interpreted as a ``B-field''. (Indeed, for a
finite abelian group, transgression provides a group isomorphism between the 
cohomology group $H^2(A,\mathbbm C^\times)$ and the group of alternating bihomomorphisms.)
\\
Again we can explicitly identify an invertible topological surface defect: the one
given by the diagonal subgroup $A_{\mathrm{diag}}\,{\subset}\, A\,{\oplus}\, A$ together with
a representative of a cohomology class in $H^2(A,\mathbbm C^\times)$, which we identify
with the associated alternating bihomomorphism $\beta$. The corresponding braided
equivalence can be computed to act as
  $$
  \begin{array}{rll}
  A\oplus A^* &\!\!\longrightarrow \!\!& A\oplus A^* \\[4pt]
  (g,\chi)&\!\! \longmapsto \!\!& (g,\chi+\beta(g,-)) \,.
  \end{array}
  $$
Thus this type of automorphism shifts the character by the contraction of the B-field 
$\beta$ with the group element.
\\[4pt]
(3)\, \emph{Electric-magnetic dualities.}
\\[2pt]
It is finally expected that electric-magnetic dualities form a class
of symmetries of Dijk\-graaf-Witten theories (see e.g.\ \cite{BCKA13} for
a discussion of such symmetries for Turaev-Viro theories based on finite-dimensional
semisimple Hopf algebras).
\\
To present a simple example, assume that $A$ is a cyclic group and fix a 
group isomorphism $\delta\colon A\,{\to}\, A^*$. Then there is a natural braided
equivalence
  $$
  \begin{array}{rll}
  A\oplus A^* &\!\!\longrightarrow \!\!& A\oplus A^*\\[4pt]
  (g,\chi)&\!\! \longmapsto \!\!& (\delta^{-1}(\chi),\delta(g)) \,.
  \end{array}
  $$
This can again be described by an invertible surface defect, namely the one given
by the diagonal subgroup $A_{\mathrm{diag}}\,{\subset}\, A\,{\oplus}\, A$ and
the alternating bihomomorphism $\beta$ defined by
  $$
  \beta(g_1,g_2) := \frac{\delta(g_1)(g_2)}{\delta(g_2)(g_1)}
  \quad\text{ for }~ g_1,g_2\,{\in}\, A\,.
  $$

A careful investigation \cite{FPSV14} of the structure of the finite
group $\mathrm{O}_q(A\oplus A^*)$ shows that in fact the three types of symmetries 
listed above generate this group, i.e.\ they already provide \emph{all} 
symmetries of the theory.


\section{Conclusions}

Here is a brief summary of crucial insights presented in this contribution:

\begin{itemize}
\item
Topological defects are important structures in quantum field theories.
\\[-8pt]~

\item
By general principles they describe symmetries and dualities of quantum field theories
in such a way that the action of these symmetries and dualities on all different
structures of a quantum field theory becomes apparent. This includes the action on
field insertions as well as the action on boundary conditions and defects of various
codimensions.
\\[-8pt]~

\item
Topological codimension-1 defects and boundaries provide a natural setting for
studying relative field theories. This includes in particular anomalous field theories and
the TFT construction of RCFT correlators.
\\[-8pt]~

\item
Applications of topological defects virtually concern all fields of physics
to which quantum field theory is applied.
\end{itemize}

\noindent
There are several classes of quantum field theories in which topological defects can be
treated quite explicitly. Besides two-dimensional rational conformal field theories, these
include three-dimensional topological field theories of Reshetikhin-Turaev type. 
We have shown that a natural class of defects and boundary conditions in such theories
is obstructed, with an obstruction taking values in the Witt group of non-degenerate braided
fusion categories. We have also
demonstrated, for the subclass of Dijkgraaf-Witten theories, how defects and boundary conditions
can be formulated in a gauge-theoretic setting using (generalizations of) relative bundles.
This provides a natural embedding of structures of categorified representation theory, in
particular of the theory of module and bimodule categories over monoidal categories, into the
more comprehensive setting of (extended) topological field theory.


\bibliographystyle{amsalpha}

\end{document}